\documentclass[aps,groupedaddress,onecolumn]{revtex4}
\usepackage{graphicx} 
\usepackage{dcolumn} 
\usepackage{bm} 
\usepackage{amsmath}
\usepackage{amssymb}
\usepackage{placeins}
\usepackage[usenames]{color} 

\begin{document}

\title{What rotation rate maximizes heat transport in rotating Rayleigh-B\'{e}nard convection with Prandtl number larger than one?}

\author{Yantao Yang}
\affiliation{SKLTCS and Department of Mechanics and Engineering Science, BIC-ESAT,
 College of Engineering, and Institute of Ocean Research, 
 Peking University, Beijing 100871, China}

\author{Roberto Verzicco}
\affiliation{Physics of Fluids Group, Department of Science and Technology, MESA+ Institute, Max Planck Center Twente for Complex Fluid Dynamics, and J. M. Burgers Center for Fluid Dynamics, University of Twente, 7500 AE Enschede, The Netherlands}
\affiliation{Dipartimento di Ingegneria Industriale, University of Rome ``Tor Vergata'', Via del Politecnico 1, Roma 00133, Italy}
\affiliation{Gran Sasso Science Institute - Viale F. Crispi 7, 67100 L'Aquila, Italy}

\author{Detlef Lohse}
\affiliation{Physics of Fluids Group, Department of Science and Technology, MESA+ Institute, Max Planck Center Twente for Complex Fluid Dynamics, and J. M. Burgers Center for Fluid Dynamics, University of Twente, 7500 AE Enschede, The Netherlands}

\author{Richard J.A.M. Stevens}
\affiliation{Physics of Fluids Group, Department of Science and Technology, MESA+ Institute, Max Planck Center Twente for Complex Fluid Dynamics, and J. M. Burgers Center for Fluid Dynamics, University of Twente, 7500 AE Enschede, The Netherlands}

\date{\today}

\begin{abstract}
The heat transfer and flow structure in rotating Rayleigh-B\'enard convection are strongly influenced by the Rayleigh ($Ra$), Prandtl ($Pr$), and Rossby ($Ro$) number. For $Pr\gtrsim 1$ and intermediate rotation rates, the heat transfer is increased compared to the non-rotating case. We find that the regime of increased heat transfer is subdivided into a low and a high $Ra$ number regime. For $Ra\lesssim 5\times10^8$ the heat transfer at a given $Ra$ and $Pr$ is highest at an optimal rotation rate, at which the thickness of the viscous and thermal boundary layer is about equal. From the scaling relations of the thermal and viscous boundary layer thicknesses, we derive that the optimal rotation rate scales as $1/Ro_\mathrm{opt} \approx 0.12 Pr^{1/2}Ra^{1/6}$. In the low $Ra$ regime the heat transfer is similar in a periodic domain and cylindrical cells with different aspect ratios, i.e.\ the ratio of diameter to height. This is consistent with the view that the vertically aligned vortices are the dominant flow structure. For $Ra\gtrsim 5\times10^8$ the above scaling for the optimal rotation rate does not hold anymore. It turns out that in the high $Ra$ regime, the flow structures at the optimal rotation rate are very different than for lower $Ra$. Surprisingly, the heat transfer in the high $Ra$ regime differs significantly for a periodic domain and cylindrical cells with different aspect ratios, which originates from the sidewall boundary layer dynamics and the corresponding secondary circulation.
\end{abstract}

\maketitle

\section{Introduction}
Since the seminal experiments by Rossby \cite{ros69}, rotating Rayleigh-B\'enard convection \cite{ahl09,loh10}, i.e.\ the buoyancy-driven flow of a fluid layer heated from below and cooled from above and rotating about the central vertical axis, has been a model system to study the influence of rotation on heat transfer \cite{ste13b}. Improving our understanding of the influence of rotation on heat transport is crucial from a technological point of view to better understand important industrial processes \cite{joh98}. It is also essential to better understand the effect of rotation on relevant natural processes such as the thermohaline circulation in the oceans \cite{rah00}, atmospheric flows \cite{har01}, trade winds \cite{had35}, zonal flows in planets like Jupiter \cite{ing90}, and the effect of rotation on reversals of the Earth's magnetic field \cite{gla95}.

With the development of experimental techniques \cite{zho93,jul96,liu97,vor02,kun08d,kin09,liu09,nie10,zho10c,wei10,wei11,wei11b,liu11,kin12b,wei15,wei16} and simulations \cite{kun08d,zho09,ste09,sch09,kun10,kun10b,ste10a,ste10b,wei10,sch10b,kun11,ste11b,ste12b,kin12b,kin13,ste14e,hor15,kun16,hor18}, significant progress on our understanding of rotating convection has been realized. Rotating Rayleigh-B\'enard convection is characterized by several major flow transitions, which strongly influence the heat transport and flow structures \cite{ste13b}. As rotation is known to have a stabilizing effect on fluid flow a particular intriguing phenomenon is the observation of a substantial heat transport enhancement at moderate rotation rates. The mechanism responsible for this heat transport enhancement is Ekman pumping \cite{ros69,jul96b,vor02,kun08d,kin09,zho09,ste09,ste10a,ste10b,ste11b}, i.e.\ due to the rotation, rising or falling plumes of hot or cold fluid are stretched into vertically aligned vortices that suck fluid out of the thermal boundary layers adjacent to the bottom and top plates. A better understanding of the transitions between these regimes, and the physics that dictates them, is of paramount importance to understand the convection phenomena described above.

Zhong {\it et al.}\ \cite{zho09} and Stevens {\it et al.}\ \cite{ste09,ste10a} found in experiments and direct numerical simulations that the Rayleigh number $Ra$ and the Prandtl number $Pr$, to be defined explicitly below, strongly influence the Ekman pumping process. As a result, the heat transfer enhancement compared to the non-rotating case strongly depends on these control parameters. They found that at a fixed non-dimensional rotation rate of $1/Ro$, also to be defined below, the heat transport enhancement is highest for {\it intermediate} $Pr$. For lower $Pr$, the efficiency of Ekman pumping is limited by the heat diffusing out of the vertically aligned vortices due to the high thermal diffusivity. For higher $Pr$ the thermal boundary layer is much thinner than the viscous boundary layer, where the base of the vortices forms, and this limits the amount of hot fluid that enters the vortices at the base. Furthermore, the effect of Ekman pumping reduces with increasing $Ra$. The reason is the increase of the turbulent diffusion, which limits the ability of the vertically aligned vortices to transport heat effectively. 

In this work, we study the heat transfer and flow structures in rotating Rayleigh-B\'enard convection as a function of the main control parameters of the system. These control parameters are the Rayleigh number $Ra = \beta g \Delta L^3/(\kappa\nu)$, where $\beta$ is the thermal expansion coefficient, $g$ the gravitational acceleration, $L$ and $\Delta$ the distance and temperature difference between the bottom and top plates, respectively, and the Prandtl number $Pr = \nu /\kappa$, where $\nu$ and $\kappa$ are the kinematic viscosity and the thermal diffusivity, respectively. The rotation rate $\Omega$ is non-dimensionalized in the form of the Rossby number $Ro = \sqrt{\beta g \Delta /L}/(2\Omega)$. As $Ro$ varies as an inverse rotation rate, we indicate the non-dimensional rotation rate as $1/Ro$ in this work. Alternatively, the strength of the rotation can be characterized by the Ekman number $Ek = \nu/L^2\Omega$ or the Taylor number $Ta=(2/Ek)^2=Ra/(Pr Ro^2)$ The heat transfer is given by the Nusselt number $Nu$, which is the ratio between the convective and conductive heat flux. The Reynolds number $Re$ measures the strength of the flow. 

The effects of the specific geometry of the domain will also be investigated in the current study. We run simulations with horizontally periodic Cartesian domain and cylindrical cells with two different aspect ratios for a wide range of $Pr$ and $Ra$. Experimental data from Refs.\ \cite{zho09,zho10c,wei11,ste12b} are also included for comparison. It has long been recognized that domain geometry has strong influences on flow properties for (rotating) Rayleigh-B\'enard convection, e.g.\ see \cite{day01,son10,liu11}. The aspect ratio of the cylinder can alter the formation of secondary flows in the Ekman and Stewartson layers~\cite{kun16,kun13}. Recent studies further reveal that the boundary zonal flow close to the sidewall in a slender cylinder exhibits very rich structures and dynamics for both momentum and temperature fields~\cite{wit20,zha20}. Here we will focus on the heat transfer enhancement and its dependence on the domain geometry, i.e., periodic domains and cylinders with different aspect ratios. 

The remainder of the manuscript is organized as follows. In section \ref{section_method} we describe the simulation methods used in this study. To study the transition from the low to the high $Ra$ regime, we performed simulations in a periodic and a cylindrical domain. In section \ref{section_overview} we discuss the observation of the low and the high $Ra$ regime based on the heat transfer data obtained from simulations and corresponding experimental data published in literature \cite{zho09}. In section \ref{section_results} we discuss the main flow features in the different regimes. The conclusions and an outlook to future work are given in section \ref{section_conclusions}.

\begin{figure}[t!]
\centering
\includegraphics[width=0.45\textwidth]{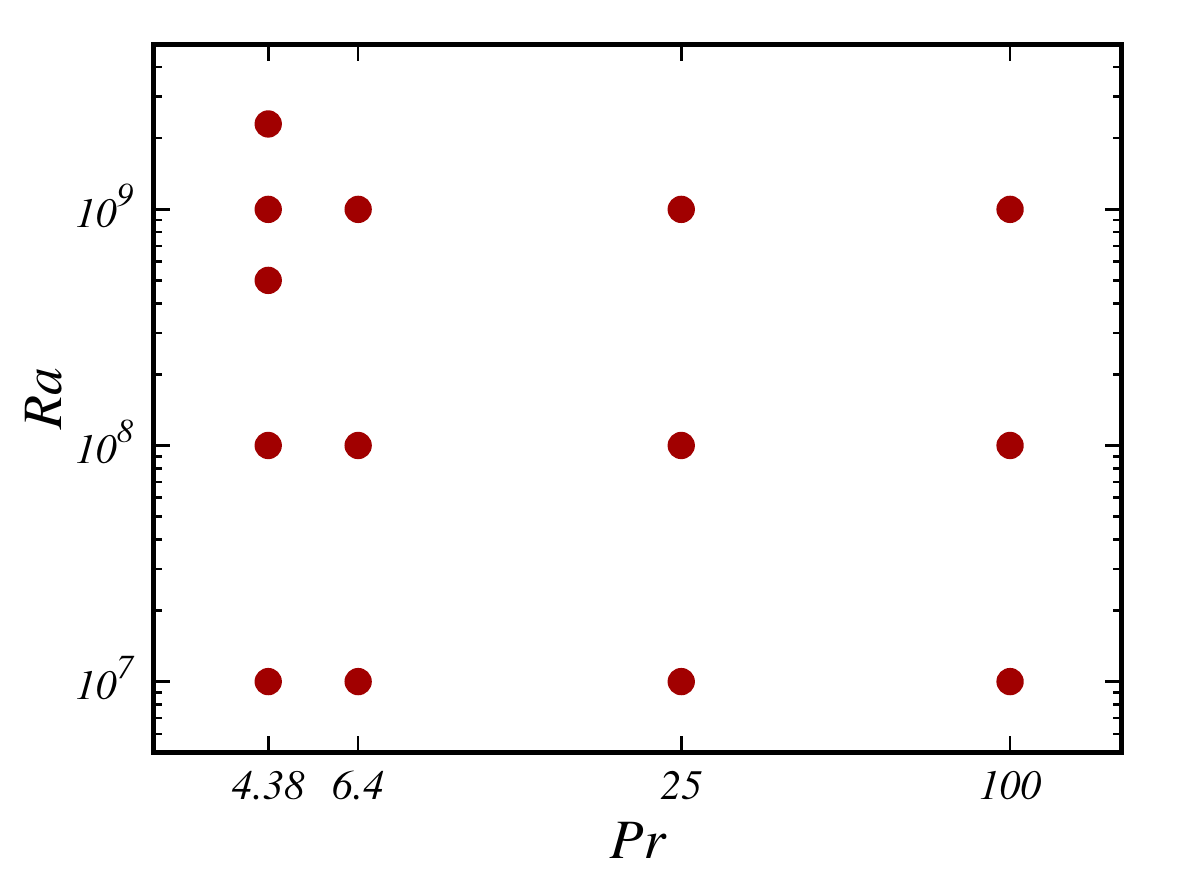}
\caption{Simulated $Ra$ and $Pr$ combinations in a horizontally periodic domain. For each point a series of simulations for various $1/Ro$ is performed. Further details on the simulations can be found in the appendix.}
\label{fig_pr-ra}
\end{figure}

\section{method} \label{section_method}

We perform direct numerical simulations by solving the three-dimensional Navier-Stokes equations within the Boussinesq approximation. We consider two different system geometries, namely a cylindrical and a Cartesian horizontally periodic domain. We use our in-house code, which has been extensively validated for Rayleigh-B\'{e}nard turbulence. The code employs a second-order finite-difference scheme with a fractional-time-step step method. For simulations in the cylinder domain, the code is the same as in our previous studies, see e.g.\ Refs~\cite{ste10a,ste10b}. For simulations in the periodic domain at high $Ra$ and $Pr$, the multiple-resolution method for scalar turbulence is used to improve the computational efficiency~\cite{ost15}. In this method the momentum equations are solved on a base mesh while the scalar field is solved on a refined mesh.

Constant temperature and no-slip boundary conditions at the bottom and top plates are employed. For the cylindrical domain, we use an adiabatic sidewall and we consider $Pr=4.38$ in a $\Gamma=1$ cylindrical cell up to $Ra=1.8\times10^{10}$. These simulations show excellent agreement with previous measurements performed by Zhong and Ahlers \cite{zho09}. From previous work we have datasets for $\Gamma=1/2$ (for $Pr=4.38$ and for $Ra$ up to $Ra=4.52\times10^{9}$) and $\Gamma=1$ (for $Ra=10^8$ and various $Pr$) available. For the Cartesian periodic domain we performed simulations for various $Ra$ and $Pr$ combinations as indicated in figure \ref{fig_pr-ra}. For each pair of $Ra$ and $Pr$ we gradually increased the rotation rate $1/Ro$ from zero to a strong enough rotation to ensure that the heat transfer is lower than for the non-rotating cases. We used a horizontal domain width that is much larger than the typical size of flow structures in the bulk to ensure that the periodic boundary condition is appropriate. As the horizontal length scale of the vertically aligned vortices decreases with increasing $Ra$ and $Pr$, we can reduce the domain width accordingly to save computing resources. Moreover, for most rotating cases the domain size is more than 10 times larger than the most unstable wavelength for convection instability which scales asymptotically as $L_c=4.82Ek^{1/3}$~\cite{kun16}. It has been shown that such domain size for rotating RB is enough to assure the convergence of the Nusselt number~\cite{kun16}. Further details about the simulations are summarized in the appendices.

\section{Overview} \label{section_overview}

\begin{figure}
 \centering
 \includegraphics[width=0.7\textwidth]{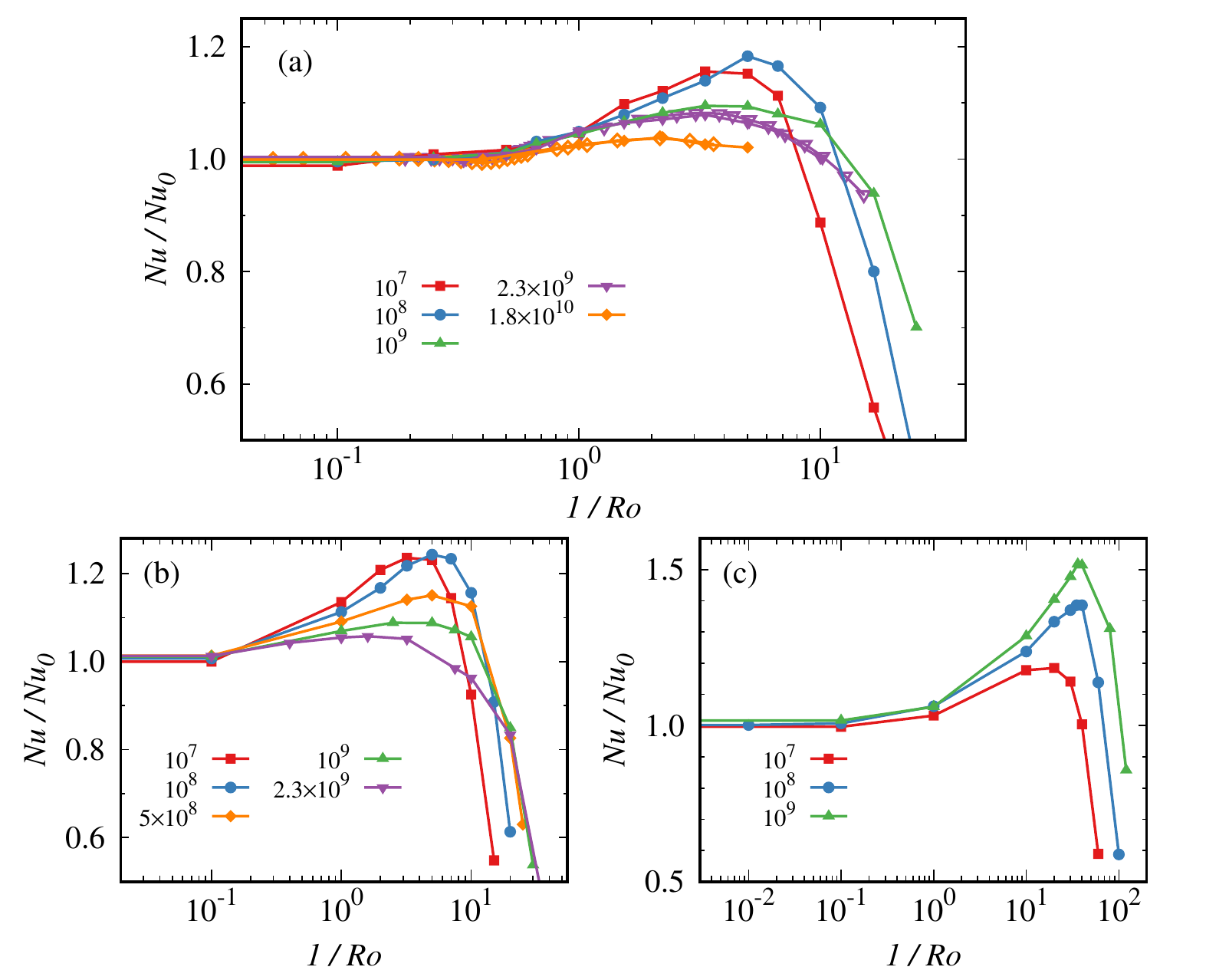}
 \caption{$Nu/Nu_0$ versus $1/Ro$ for different $Ra$ as indicated in the legend. (a) Experimental (open symbols; \cite{zho09}) and simulation (solid symbols) results for $Pr=4.38$ and in a $\Gamma=1$ cylinder. Panel (b) and (c) show simulation results for $Pr=4.38$ and $Pr=100$, respectively, obtained in a periodic domain.}
 \label{fig_nu-ro}
\end{figure}

\begin{figure}
 \centering
 \includegraphics[width=0.7\textwidth]{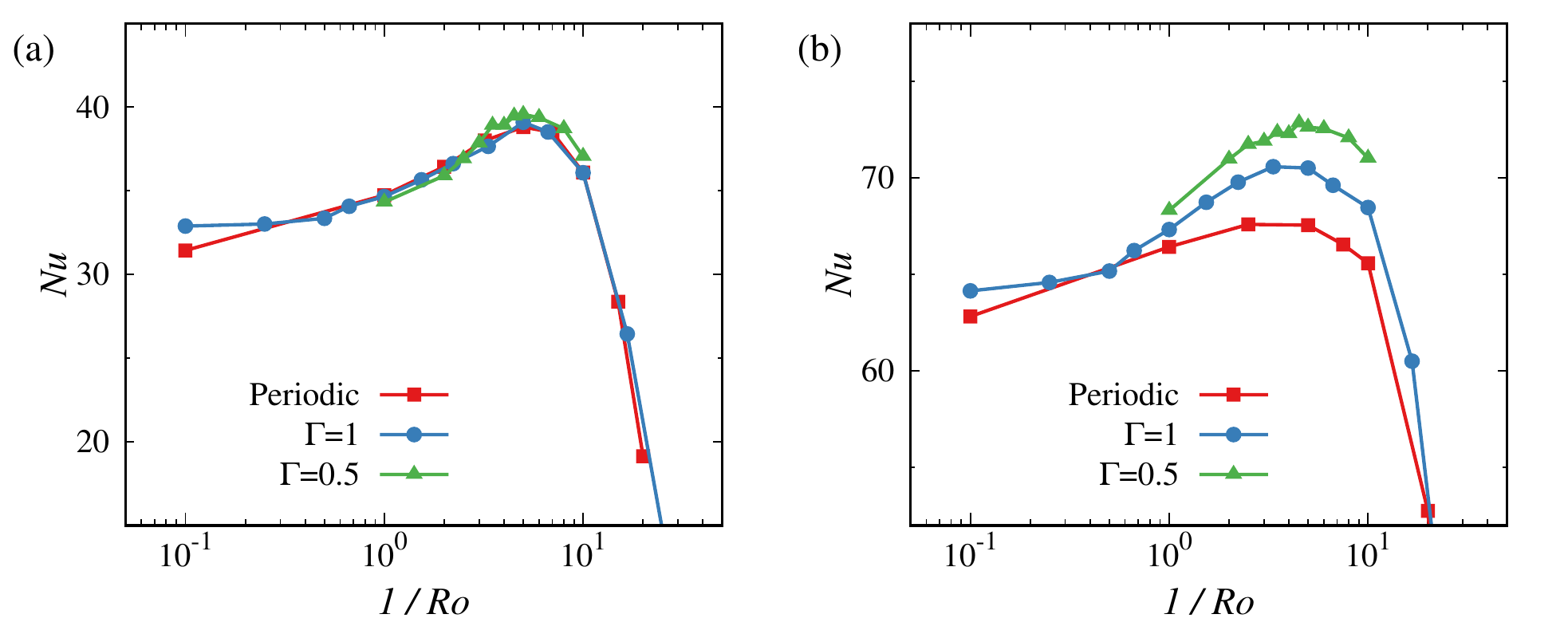}
 \caption{Comparison of $Nu$ obtained in a periodic and in cylindrical domains with $\Gamma=1/2$ and $\Gamma=1$ for $Pr=4.38$ at (a) $Ra=10^8$ and (b) $Ra=10^9$. The $\Gamma=1/2$ data are from Ref.\ \cite{ste12b}.}
\label{fig_nudomain}
\end{figure}

Figure \ref{fig_nu-ro} shows the heat transfer enhancement with respect to the non-rotating case, i.e.\ $Nu/Nu_0$, versus the non-dimensional rotation rate $1/Ro$ for various cases. Figure \ref{fig_nu-ro}a shows that we obtain excellent agreement between simulations and the experiments performed by Zhong and Ahlers \cite{zho09}. Similarly to previous studies, we find that the heat transfer first increases for moderate rotation rates before it quickly decreases for strong rotation rates. The optimal rotation rate is defined as the rotation rate $1/Ro_\mathrm{opt}$ for which the heat transfer ($Nu$) for that $Ra$, $Pr$, and aspect ratio $\Gamma$ is maximal. For $Ra\lesssim 5 \times10^8$ and $Pr=4.38$ the optimal rotation rate increases with increasing $Ra$ before it rapidly decreases with increasing $Ra$ when $Ra\gtrsim 5 \times10^8$. This indicates that the flow dynamics that determine the optimal rotation rate are different in the low and the high $Ra$ regime. Figures \ref{fig_nu-ro}b and c show results from the periodic domain simulations for $Pr=4.38$ and $Pr=100$ and various $Ra$. A direct comparison of the heat transfer data for $Pr=4.38$ in figure \ref{fig_nudomain} reveals that for $Ra=10^8$ the heat transfer behaves almost identically in a periodic domain and cylinders with different aspect ratios, while there are surprisingly significant differences for $Ra=10^9$. 

\begin{figure}
 \centering
 \includegraphics[width=0.84\textwidth]{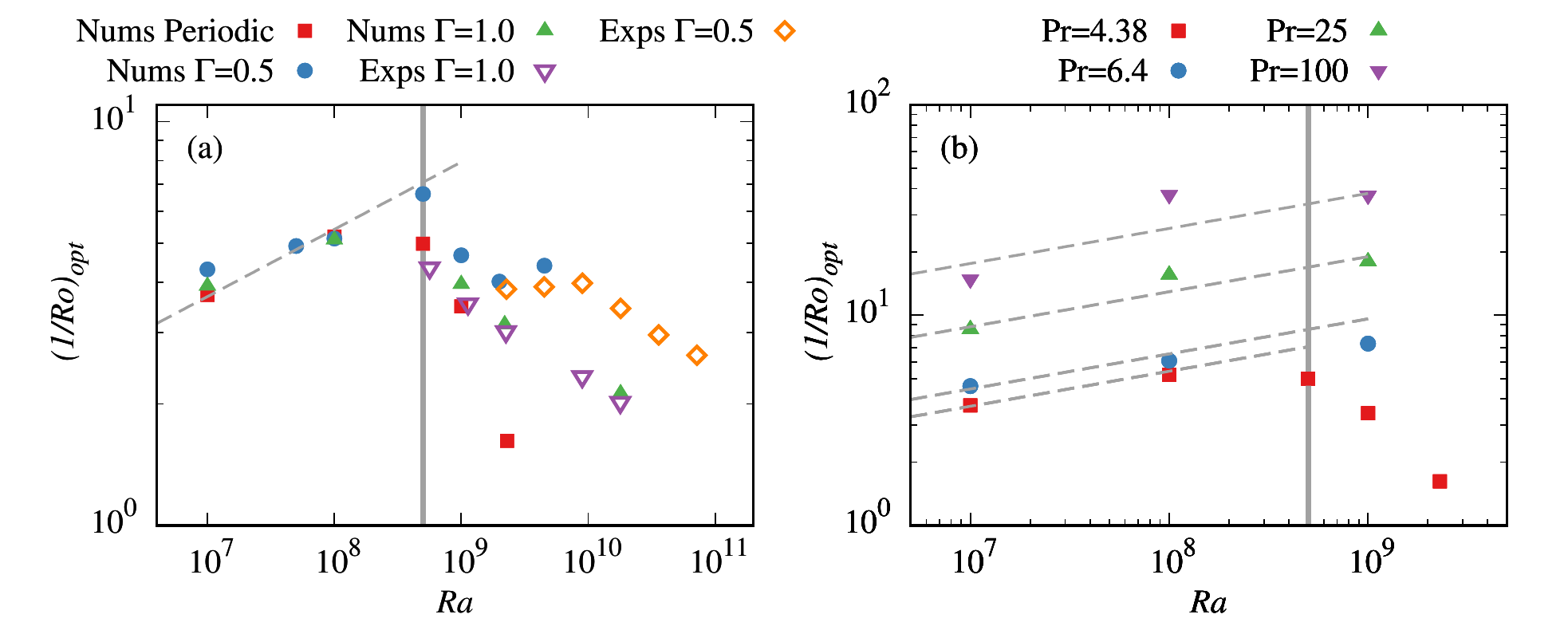}
 \caption{The optimal rotation rate for different $Ra$ for (a) $Pr=4.38$ and different geometries, i.e.\ a periodic domain and cylindrical domains with $\Gamma=0.5$ and $\Gamma=1$, and (b) for different $Pr$ in a periodic domain. The vertical line marks the transition from the low to high $Ra$ regime. The dashed lines indicate the boundary layer scaling law $1/Ro \approx 0.12 Pr^{1/2}Ra^{1/6}$. The experimental results are from Refs.\ \cite{zho09,zho10c,wei11} and the $\Gamma=1/2$ simulation data are from Ref.\ \cite{ste12b}}
 \label{fig_peak}
\end{figure}

To reveal the transition between the low and the high $Ra$ regime more clearly, we plot the optimal rotation rate as a function of $Ra$ and $Pr$ for all available cases in figure \ref{fig_peak}. To determine the optimal rotation rate from simulation data, we perform a second-order polynomial fit around the rotation rate at which the heat transport is highest and determine the optimal rotation rate using that polynomial fit. For experimental results, we first determine the largest heat transfer enhancement $Nu_\mathrm{max}-Nu_0$, where $Nu_\mathrm{max} $ is the highest heat flux obtained for fixed $Pr$ and $Ra$ and different $Ro$. Subsequently, we determine the optimal rotation rate from a polynomial fit to all data points for which $Nu-Nu_0 > 0.5(Nu_\mathrm{max} -Nu_0)$. From figure \ref{fig_peak}(a) the existence of a low and a high $Ra$ regime and their different features are immediately apparent. For $Pr=4.38$ and $Ra \lesssim 5\times10^8$ the optimal rotation rate is well described by the following scaling
\begin{equation}\label{eq:peakscaling}
 (1/Ro)_\mathrm{opt} \approx 0.12 Pr^{1/2} Ra^{1/6}.
\end{equation}
The derivation of this scaling law will be given in section \ref{section_lowRa}. For $Ra \lesssim 5\times10^8$ the data for the different domain geometries is well described by the above scaling. However, we find that in contrast to the above prediction the optimal rotation rate decreases with increasing $Ra$ for $Ra \gtrsim 5\times10^8$.

Although for $Pr=4.38$ the transition to the high $Ra$ number regime is very pronounced, such a transition can not be seen clearly in the high $Pr$ number data. Studying the transition at higher $Pr$ would require more high $Ra$ number simulations. However, unfortunately, such simulations are too time-consuming to be performed. In the following discussion, for the low $Ra$ regime, we include all cases with $Ra\le 5\times10^8$ and all $Pr$'s. While for the high $Ra$ regime, we will limit ourselves to $Pr=4.38$. We explain the different behaviors of the optimal rotation rate in the low and high $Ra$ regime, and the physical mechanism behind it.

\section{Optimal heat transfer in the low and high $Ra$ regime} \label{section_results}

\subsection{The low $Ra$ regime ($Ra\leq 5\times10^8$)} \label{section_lowRa}

It has been conjectured that the heat transfer reaches a maximum when the viscous and thermal boundary layers have a similar thickness \cite{jul96,kin09,liu09,kin13,eck14,hor14,raj17}. The importance of this boundary layer transition has been recognized in several previous works. For example, King {\it et al.}\ \cite{kin12} used it to derive a scaling law to describe the transition to the geostrophic convection regime \cite{sak97,kin12,eck14,ste14e,plu16}. In rotating Rayleigh-B\'enard convection the Ekman boundary layer thickness scales as~\cite{gre90,kin09,kin12,kin12b,kin13}
\begin{equation} \label{EkmanScaling2}
 \lambda_u/L \sim 1/Ta^{1/4} \sim (1/Ro)^{-1/2}Pr^{1/4}Ra^{-1/4}.
\end{equation}
The thickness of the thermal boundary layer is related to the scaling of the $Nu$ number. For non-rotating convection the $Nu$ number as function of $Ra$ and $Pr$ is well described by the unifying theory for thermal convection \cite{gro00,gro01,ste13}. For {\it intermediate} $Pr$ and before the onset of the ultimate regime \cite{ahl09} one obtains that the thermal boundary layer thickness $\lambda_\theta$ can be {\it approximated} as
\begin{equation} \label{thermalboundary layer2}
 \lambda_{\theta}/L \sim (1/Ro)^{0}Pr^{0}Ra^{-1/3} .
\end{equation}
Obviously, this relation is a {\it simplification} that is used in the analysis \cite{kin09,kin12,kin12b,kin13}, which does not do full justice to the Rayleigh-B\'enard dynamics as for non-rotating convection and $Pr \sim 1$ the scaling exponent $\gamma$ in $Nu \sim Ra^\gamma$ depends on $Ra$ as described by the unifying theory \cite{gro00,gro01,ste13}. For {\it intermediate} $Pr$ and before the onset of the ultimate regime the effective scaling exponent typically is in the range $0.28-0.32$ \cite{ahl09}, which is close to the approximation used above, and sufficiently accurate to not affect the discussion presented below.

Combining equations \eqref{EkmanScaling2} and \eqref{thermalboundary layer2} gives that the ratio between the boundary layer thicknesses scales as
\begin{equation} \label{scaling}
\frac{\lambda_\theta}{\lambda_{u}} \sim (1/Ro)^{1/2}Pr^{-1/4}Ra^{-1/12}.
\end{equation}
This is equivalent to ${\lambda_\theta}/{\lambda_{u}} \sim RaEk^{3/2}$~\cite{kin12}. 

\begin{figure}
 \centering
 \includegraphics[width=\textwidth]{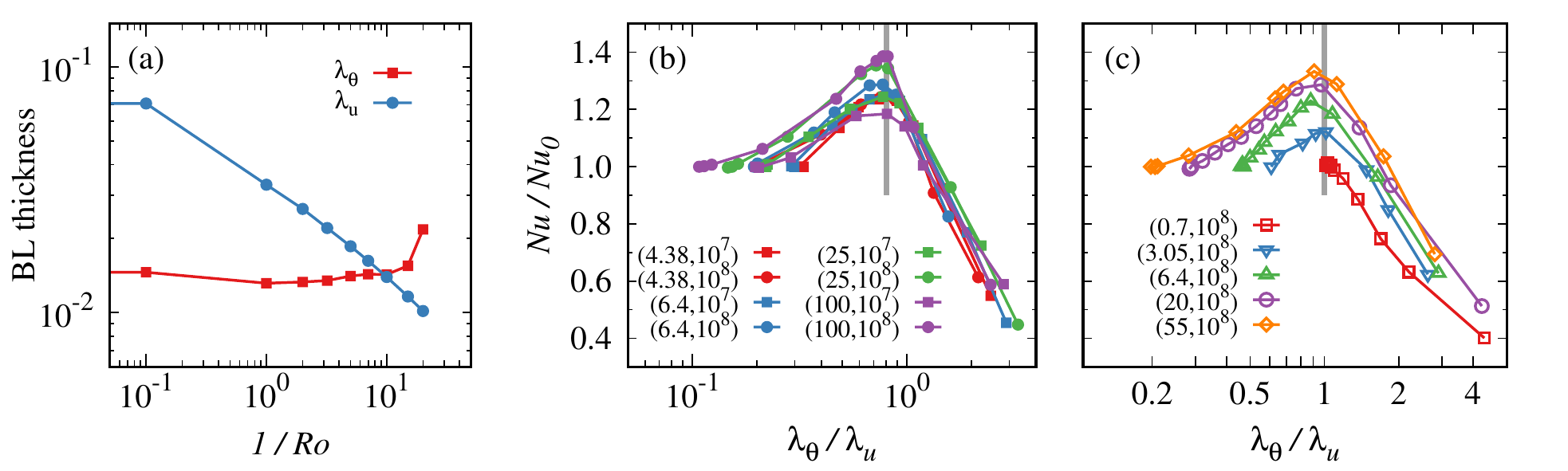}
 \caption{(a) The thermal $\lambda_\theta$ and viscous $\lambda_u$ boundary layer thickness versus $1/Ro$ for $Pr=4.38$ and $Ra=10^8$ obtained in a periodic domain. (b) $Nu/Nu_0$ as function of $\lambda_\theta / \lambda_u$ for eight different combinations of $Ra$ and $Pr$ in the low $Ra$ regime; note that the optimal rotation rate is found at $\lambda_\theta / \lambda_u \approx 0.8$, see the dashed vertical line. (c) Same as panel (b), but now for data obtained in a cylindrical domain.}
 \label{fig_bl}
\end{figure}

\begin{figure}
 \centering
 \includegraphics[width=\textwidth]{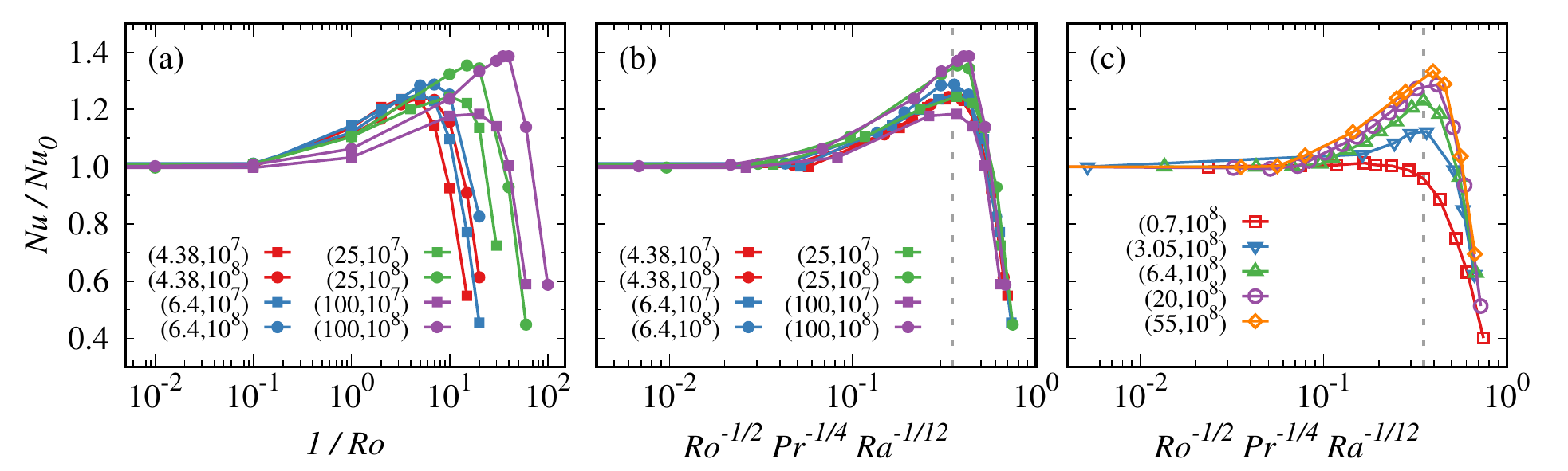}
 \caption{$Nu/Nu_0$ as function of (a) $1/Ro$ and (b) $Ro^{-1/2}Pr^{-1/4}Ra^{-1/12}$, which determines the locations of the optimal rotation rate in the low $Ra$ regime, for eight combinations of $Ra$ and $Pr$. The vertical line in (b) and (c) marks the location of the maximum heat transfer at $(1/Ro)^{1/2}Pr^{-1/4}Ra^{-1/12}\approx0.35$. (a) and (b) show the cases in periodic domain, while (c) in cylinder domain, respectively.}
\label{fig_scaling}
\end{figure}

Figure \ref{fig_peak} shows that this scaling indeed captures the position of the optimal rotation rate in the low $Ra$ regime. To further verify the assumptions in the above analysis, we show the viscous and thermal boundary layer thickness as a function of $1/Ro$ for $Ra=10^8$ and $Pr=4.38$ in figure \ref{fig_bl}a. The viscous boundary layer thickness $\lambda_u$ is determined by the peak location of the root-mean-square (rms) value of the horizontal velocities, and the thermal boundary layer thickness $\lambda_{\theta}$ by the height of the first peak location of the temperature rms profile. The figure shows that the viscous boundary layer thickness decreases monotonically with increasing rotation rate. The thermal boundary layer thickness first decreases slightly with increasing rotation before it increases rapidly with increasing rotation. The viscous and thermal boundary layer thicknesses are about equal at the optimal rotation rate of $1/Ro\approx10$. Figure \ref{fig_bl}b shows that in the low $Ra$ regime the highest heat transfer is obtained for $\lambda_{\theta}/\lambda_u\approx0.8$ for all available cases. Figure~\ref{fig_bl} shows that also in a cylindrical cell the optimal rotation rate occurs when $\lambda_{\theta}/\lambda_u\approx1.0$. For the cylindrical case, the viscous boundary layer thickness was determined as twice the height where the horizontally averaged value of $\epsilon''_u :=\langle\mathbf{u}\cdot\nabla^2\mathbf{u}\rangle_h$ is highest, as in Ref.~\cite{ste10b}. Thus in the low $Ra$ regime the optimal rotation rate is determined by the ratio of the thermal and viscous boundary layer thickness. That the scaling relation $(1/Ro) \propto Pr^{1/2} Ra^{1/6}$ is appropriate to capture the peak is further illustrated in figure \ref{fig_scaling}. Figure \ref{fig_scaling}a shows the dependence of $Nu/Nu_0$ on $1/Ro$ for different combinations of $Ra$ and $Pr$ obtained in a periodic domain. The onset of heat transfer enhancement is independent of $Ra$ and $Pr$ for $Pr>1$ \cite{wei10,ste13b}. After the onset of heat transfer enhancement, the heat transfer is strongly affected by $1/Ro$, $Pr$, and $Ra$. However, figure \ref{fig_scaling}b shows that the peak for $Nu/Nu_0$ nicely collapses for all combinations of $Ra$ and $Pr$ when the above scaling is applied. Figure \ref{fig_scaling}c shows that this scaling argument also works for data obtained in a cylindrical cell. The data in this last figure are from Ref.\ \cite{ste10} and have been supplemented here with data from additional simulations to ensure that smooth curves can be plotted for all $Pr$.

\subsection{The high $Ra$ regime ($Ra\gtrsim 5\times10^8$)}

\begin{figure}[!t]
\centering
\includegraphics[width=0.84\textwidth]{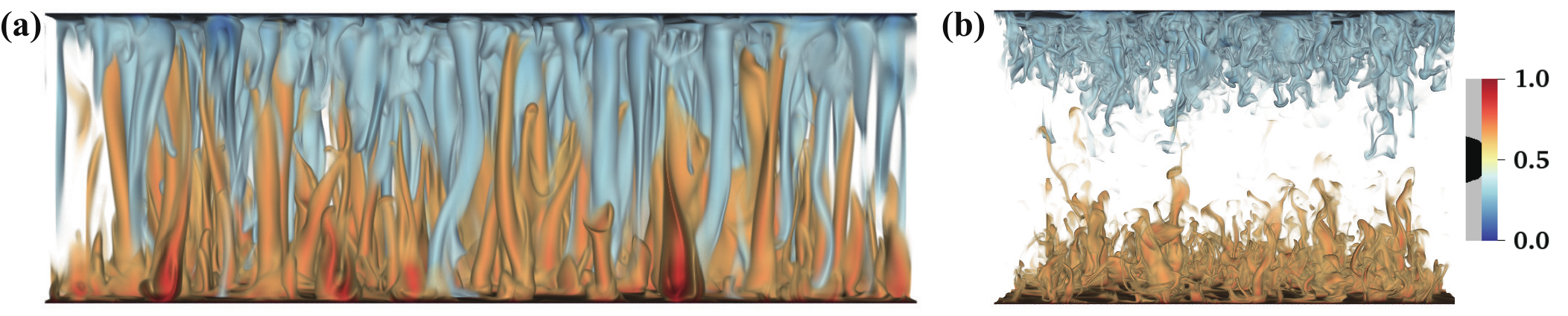}
\caption{Volume renderings of the temperature field at the optimal rotation rate for simulations with $Pr=4.38$ at (a) $Ra=10^8$ and (b) $Ra=2.3\times10^9$. The colormap in both panels is identical. The figure shows that the flow structure at the optimal rotation rate is very different in the low and high $Ra$ regime.}
\label{fig_3dfield}
\end{figure}

According to the boundary layer scaling arguments discussed above the optimal rotation rate should increase with increasing $Ra$. However, figure \ref{fig_peak} shows that for $Pr=4.38$ the high $Ra$ regime sets in around $Ra=5\times10^8$. In this high $Ra$ regime the optimal rotation rate decreases with increasing $Ra$, which shows that the boundary layer scaling argument cannot hold anymore, and the flow dynamics at the optimal rotation rate must be different. To investigate this transition, we show in figure \ref{fig_3dfield} the volume renderings of the temperature for $Ra=10^8$ and $Ra=2.3\times10^9$, which, according to figure 4, are respectively in the low and the high $Ra$ regime. The difference between the two cases is very distinct. For $Ra=10^8$ vertically aligned vortices, which extend over almost the full domain height, are visible. For $Ra=2.3\times10^9$ these vortices are much less pronounced and much shorter. Clearly, the flow structures at the optimal rotation rate are much more coherent in the low $Ra$ regime than in the high $Ra$ regime. 

\begin{figure}
 \centering
 \includegraphics[width=0.84\textwidth]{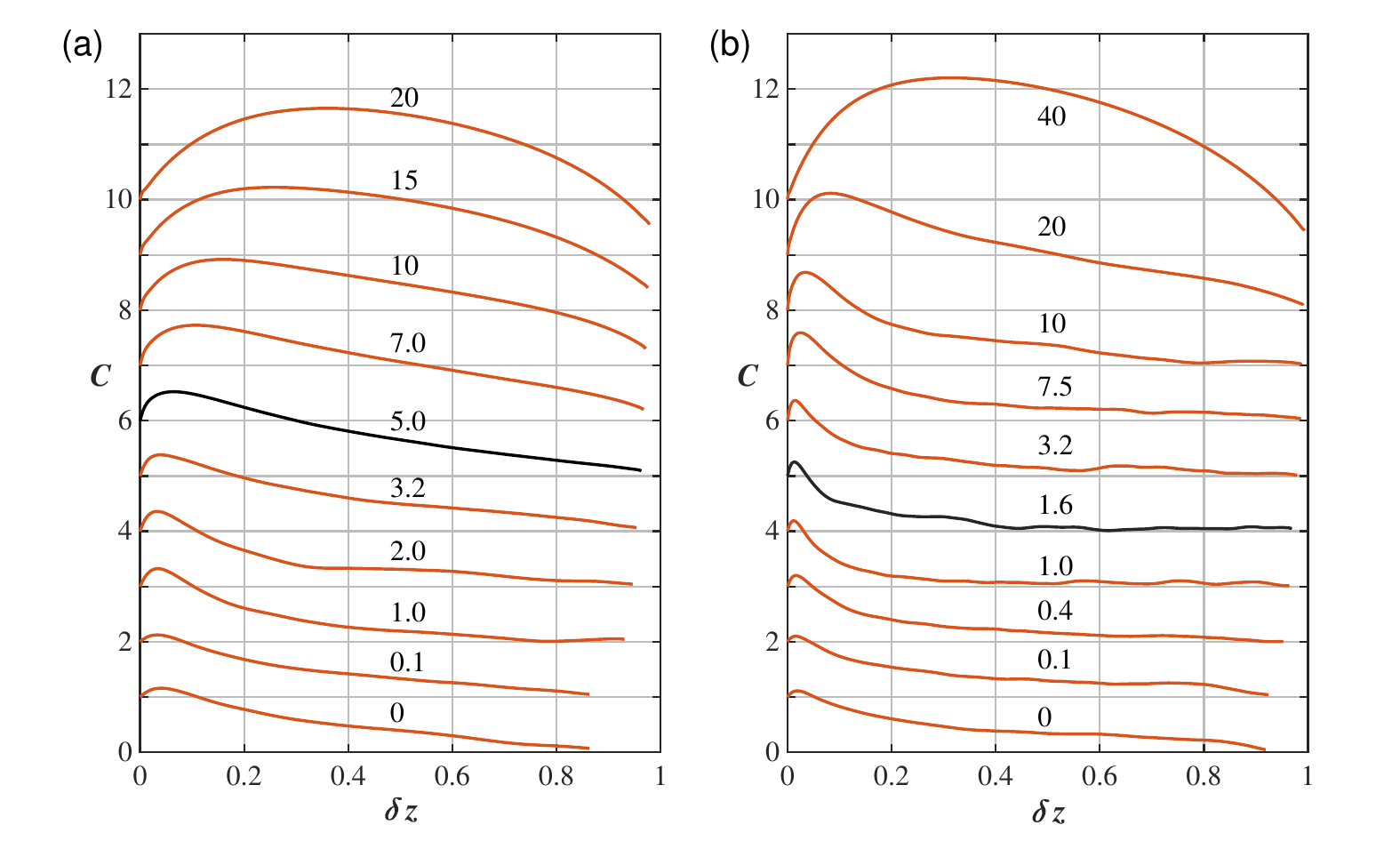}
 \caption{Correlation C as function of $\delta z$ for different rotation rates in periodic domains in the (a) low ($Ra=10^8$ and $Pr=4.38$) and the (b) high $Ra$ regime ($Ra=2.3\times10^9$ and $Pr=4.38$). The numbers indicate the value of $1/Ro$. The lines for each successive $1/Ro$ are shifted upward by $1$ for visibility. The black line indicates the optimal inverse Rossby number.}
\label{fig_corre}
\end{figure}

In an attempt to quantify the above observation we characterize the coherence of the flow structures by calculating the following cross-correlation function:
\begin{equation} \label{eq_corre}
 C(\delta z) = \frac{\langle w(x,y,\lambda_u) w(x,y,\lambda_u+\delta z)\rangle_{x,y}}
 { \langle (w(x,y,\lambda_u))^2\rangle_{x,y} },
\end{equation}
where $\langle\cdot\rangle_{x,y}$ indicates the average in horizontal direction. As we are interested in the vertically aligned vortices, we calculate the correlation using the horizontal plane at the viscous boundary layer height ($z=\lambda_u$), where the base of the vortices forms, as reference. Figure \ref{fig_corre} shows the correlation $C$ as a function of the distance from the viscous boundary layer $\delta z$ for different rotation rates in the low and high $Ra$ regime. For all cases the correlation $C$ first increases from $1$ for $\delta z=0$ to some maximum value before it decreases below $1$ further away from the boundary layer. The reason for the maximum is that the vertical velocities are higher at some distance above the viscous boundary layer height than at $z=\lambda_u$. 

\begin{figure}
 \centering
 \includegraphics[width=0.84\textwidth]{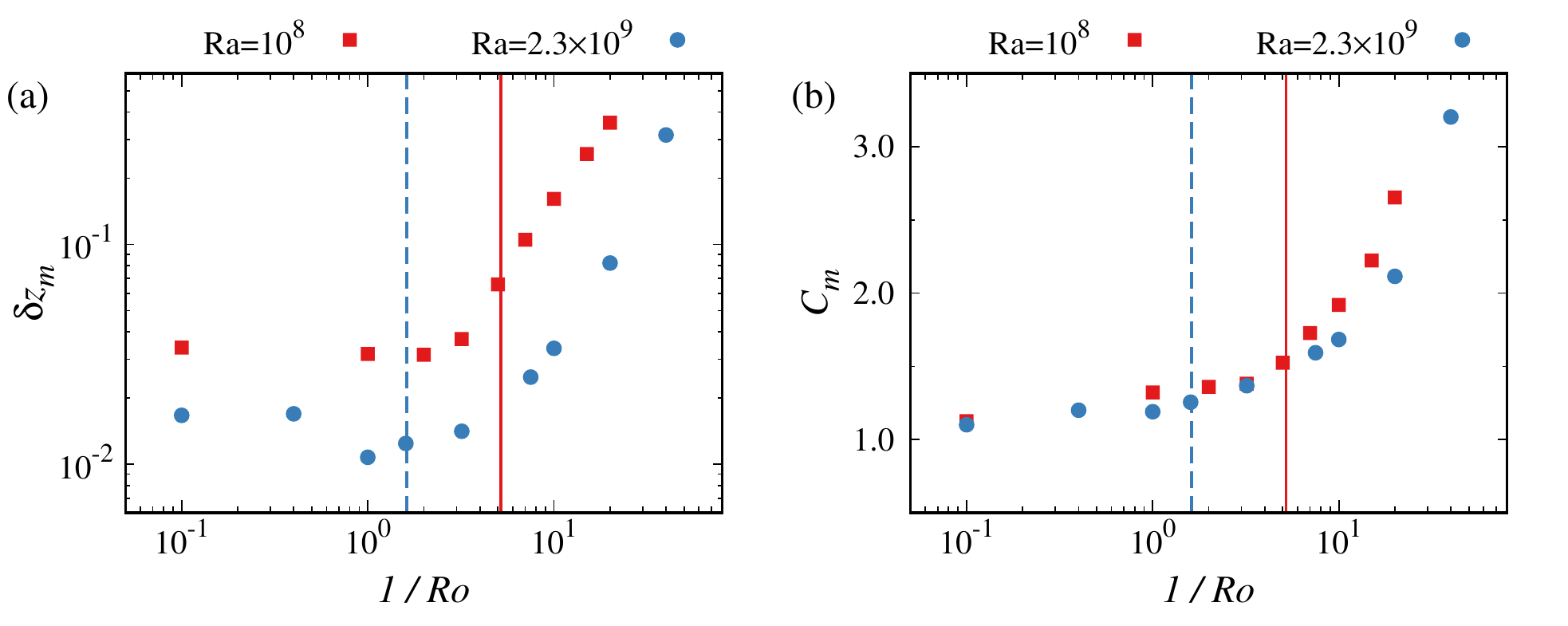}
 \caption{The (a) location $\delta z_m$ and (b) magnitude $C_m$ for the peak of the correlation function $C(\delta z)$ versus $1/Ro$ for $Pr=4.38$ and two different $Ra$ in periodic domains. The optimal rotation rate for $Ra=10^8$ and $Ra=2.3\times10^9$ is indicated by the solid and dashed line, respectively.}
\label{fig_maxcor}
\end{figure}

\begin{figure}
 \centering
 \includegraphics[width=0.84\textwidth]{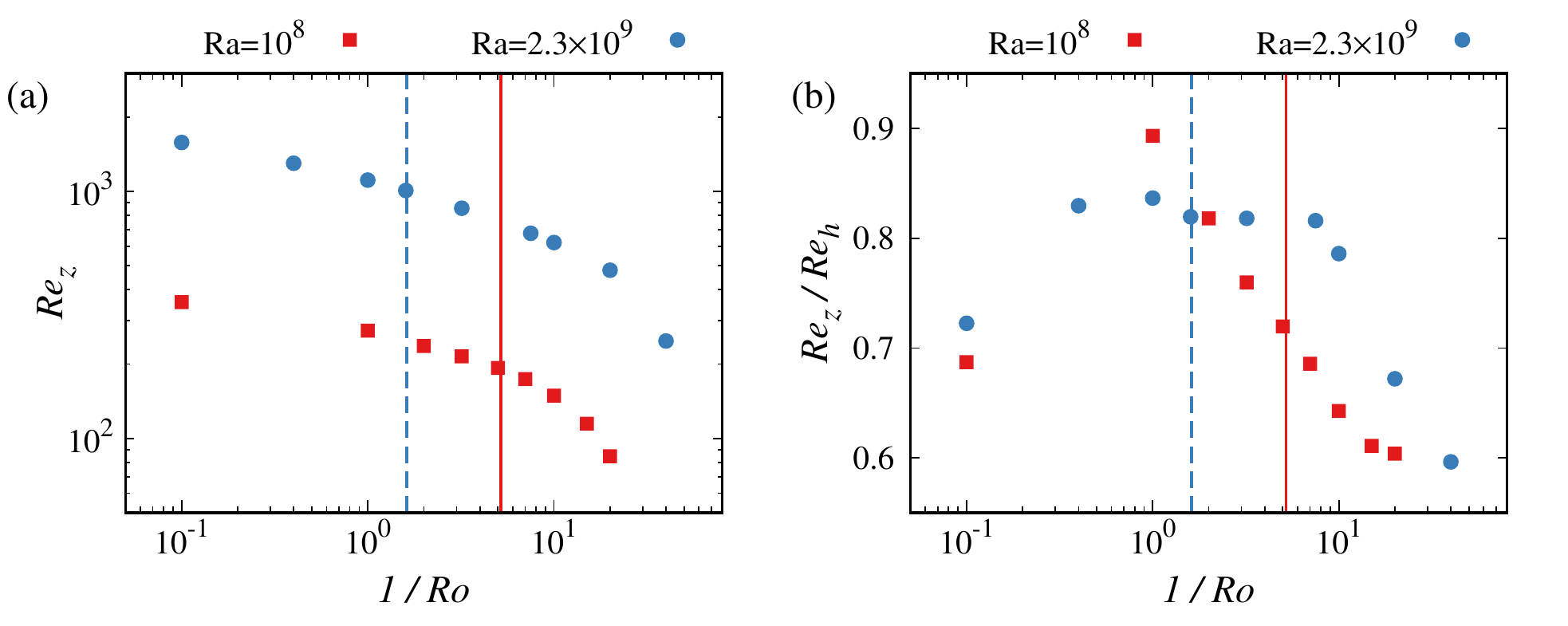}
 \caption{(a) The Reynolds number defined by the rms of the vertical velocity, and (b) the ratio of the vertical and horizontal rms velocities as function of $1/Ro$ for $Pr=4.38$ and two different $Ra$ in periodic domains. The optimal rotation rate for $Ra=10^8$ and $Ra=2.3\times10^9$ is indicated by the solid and dashed line, respectively.}
\label{fig_reuz}
\end{figure}

\begin{figure}
 \centering
 \includegraphics[width=0.84\textwidth]{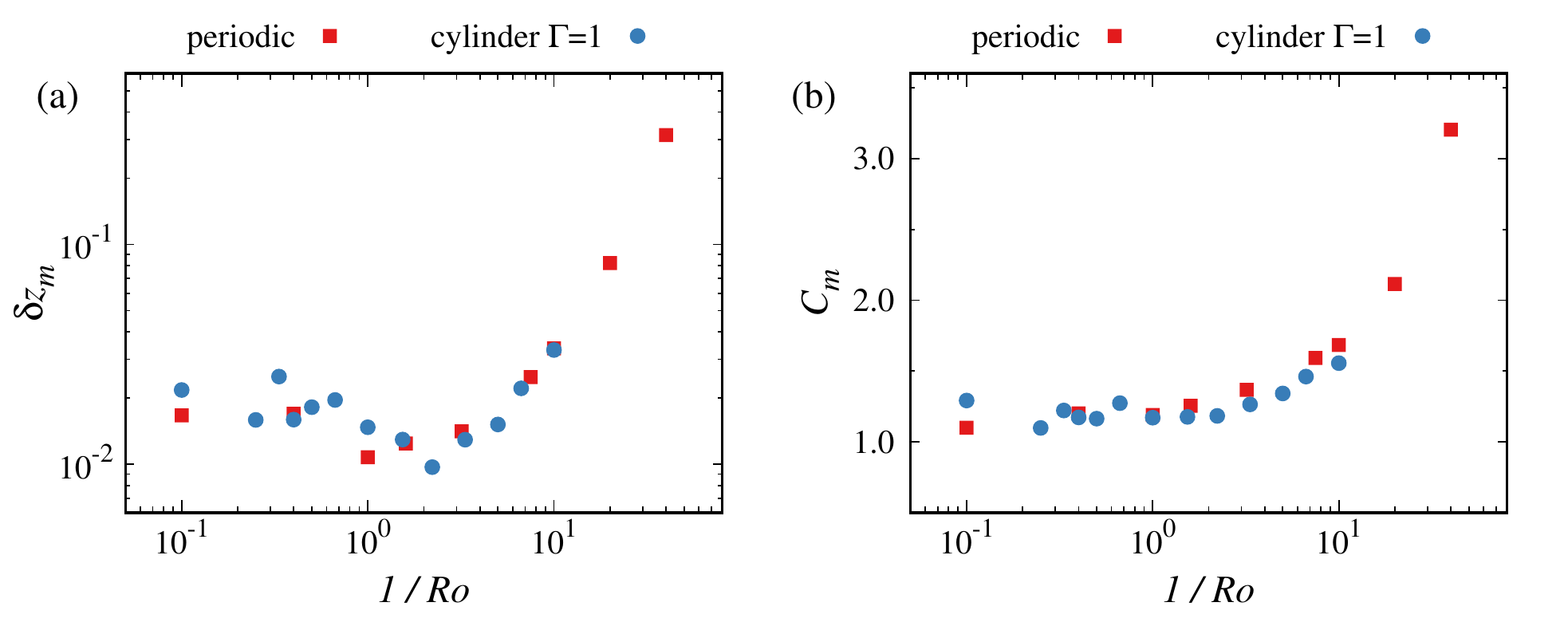}
 \caption{A comparison of (a) the location $\delta z_m$ and (b) magnitude $C_m$ for the peak of correlation function $C(\delta z)$ versus $1/Ro$ obtained from simulations performed in a periodic and cylindrical domain ($\Gamma=1$) for $Pr=4.38$ and $Ra=2.3\times10^9$. For the cylinder domain the flow data from the region close to the sidewall is excluded from the analysis.}
\label{fig_cydprd}
\end{figure}

We take the $\delta z$ value at which the maximum occurs as a measure for the height of the vertically aligned vortices and the magnitude $C_m$ of the peak of $C(\delta z)$ as a measure of the coherence of the vortices. Figure \ref{fig_maxcor} shows that the variation of $C_m$ with $1/Ro$ is similar for both $Ra$, which suggests that the flow coherence mainly depends on the rotation rate. However, the height of the plumes or vertically aligned vortices, which is indicated by $\delta z_m$, is very different in the low and the high $Ra$ regime. In the low $Ra$ regime the height of the vertically aligned vortices increases for smaller $1/Ro$ than in the high $Ra$ regime. For $Ra=2.3\times10^9$ the height at which the vertical coherence is highest is similar for the non-rotating case and the optimal rotation rate. In contrast, the vertical coherence at the optimal rotation rate is significantly higher than for the non-rotating case at $Ra=10^8$.

It is well known that strong rotation not only induces vertically aligned vortices but also suppresses the vertical fluid motion. This is illustrated in figure \ref{fig_reuz}a, which shows that the vertical Reynolds number $Re_z$, defined by the rms of vertical velocity, decreases with increasing rotation rate. The reduction of the vertical velocity between the non-rotating case and the optimal rotation rate is stronger for $Ra=10^8$ than for $Ra=2.3\times10^9$. Figure \ref{fig_reuz}b shows that, as a result, the ratio of the vertical to horizontal Reynolds number at the optimal rotation rate is larger in the high $Ra$ regime than in the low $Ra$ regime.

The above analysis reveals that in the low $Ra$ regime the maximum heat transfer is observed when there is a strong coherence in the vertically aligned vortices, and in this regime the boundary layer structure controls the optimal rotation rate. In contrast, in the high $Ra$ regime, the maximum heat transfer occurs when the vertical motion is stronger than the horizontal motion, and the ratio between the viscous and thermal boundary layer thickness does not determine the optimal rotation rate anymore.

Figure \ref{fig_peak} shows that periodic domain and cylindrical cells with different aspect ratios generate similar heat transport in the low $Ra$ regime. In contrast, it turns out that in the high $Ra$ regime the transport depends very strongly on the specific geometry of the domain. Previous studies demonstrated the importance of finite size effects in rotating Rayleigh-B\'enard by showing that the rotation rate at which heat transport sets in, and the formation of secondary flows in the Ekman and Stewartson layers \cite{kun13,kun16}, depends on the aspect ratio of the domain. However, the extreme dependence of the heat transport on the domain geometry is surprising considering the small horizontal length scale of the vertically aligned vortices that are a dominant feature of the flow. In figure \ref{fig_cydprd} we compare the results for the correlation function (equation \ref{eq_corre}) for the periodic and the cylindrical domain in an attempt to explain the origin of this difference. To eliminate the effects of the sidewall and the secondary circulation, we excluded the sidewall region $0.45<r/L<0.5$ from the analysis of data for the cylindrical cases. The figure shows that both $\delta z_m$ and $C_m$ are very similar for the cylindrical and the periodic domain over the whole range of rotation rates. Thus in the bulk region, the characteristic height and the coherence of flow structures are hardly affected by the sidewall. Hence, we conclude that the difference in the heat transfer is strongly related to the sidewall boundary layer or an effect of the secondary flow circulation.

\section{Conclusions} \label{section_conclusions}

To summarize, we systematically studied the heat flux enhancement in rotating Rayleigh-B\'enard convection for a wide range of control parameters. Based on the available data, it becomes clear that there is a low and a high $Ra$ regime in rotating Rayleigh-B\'enard. In the low $Ra$ regime, the bulk is dominated by long vertically aligned vortices, and due to the strong vertical coherence a pronounced heat transport enhancement compared to the non-rotating case is observed. The optimal rotation rate occurs when the viscous and thermal boundary layer thickness are about equal. According to this argument the optimal rotation rate scale as $1/Ro_\mathrm{opt} \approx 0.12 Pr^{1/2}Ra^{1/6}$, which is equivalent to $Ek_\mathrm{opt} \approx 8.16 Ra^{2/3}$, where the numerical values have been determined by fitting experimental and simulation data. 

In the high $Ra$ regime, the optimal rotation rate decreases with increasing $Ra$. This means that the trend is opposite to the one observed in the low $Ra$ regime, which implies that the optimal rotation rate is not obtained when the viscous and thermal boundary layer thickness are similar. In the low $Ra$ regime the flow structure at the optimal rotation rate is characterized by pronounced vertically aligned vortices. However, this is not the case in the high $Ra$ regime. Instead, we find that in the high $Ra$ regime the ratio of the vertical to horizontal velocity is much larger at the optimal rotation rate than in the low $Ra$ regime. Furthermore, we demonstrate that the domain geometry has a surprisingly pronounced influence on heat transport for higher $Ra$. Our current analyses suggest that the sidewall boundary layer in the cylindrical domain is responsible for the difference, since the flow structure in the bulk show almost the same behavior for the periodic domain and the aspect ratio $1$ cylinder. Our analysis shows that the flow structure in the bulk is almost the same in a periodic domain and the central region of an aspect ratio $1$ cylinder. This suggests that the sidewall Stewartson boundary layers cause the difference between the simulations in the periodic and cylindrical domain.

Finally, although we discuss various aspects for the transition from the low $Ra$ to the high $Ra$ regime in rotating Rayleigh-B\'enard at $Pr=4.38$, many questions on this new transition remain. For example, the differences between the low and high $Ra$ regimes are not fully explored, especially for high $Pr$. While the data and a simplified analysis suggest that the maximum heat transfer in the low $Ra$ number regime is obtained when the thermal and viscous boundary layer thickness is equal, it is not clear what physical mechanism determines the optimal rotation rate in the high $Ra$ number regime. Another aspect that needs further clarification is the role of the sidewall Stewartson boundary layers. It is namely unclear why the heat transport does not depend on the system geometry in the low $Ra$ number regime, while there is a strong geometry dependence in the high $Ra$ number regime. These and other aspects need further investigation to understand better the flow dynamics in the newly discovered high $Ra$ number rotating Rayleigh-B\'enard regime.

\bigskip

{\it Acknowledgements:} We appreciate the valuable comments from the referees. This study is supported by the Netherlands Center for Multiscale Catalytic Energy Conversion (MCEC), an NWO Gravitation program funded by the Ministry of Education, Culture and Science of the government of the Netherlands. Y. Yang acknowledges the partial support from the Major Research Plan of National Nature and Science Foundation of China for Turbulent Structures under the Grants 91852107 and 91752202. We also acknowledge the financial support from ERC (the European Research Council) Starting Grant No. 804283 UltimateRB. The authors gratefully acknowledge the Gauss Centre for Supercomputing e.V. (www.gauss-centre.eu) for funding this project by providing computing time on the GCS Supercomputer SuperMUC-NG at Leibniz Supercomputing Centre (www.lrz.de). Part of the work was carried out on the national e-infrastructure of SURFsara, a subsidiary of SURF cooperation, the collaborative ICT organization for Dutch education and research. We also acknowledge PRACE for awarding us access to MareNostrum IV based in Spain at the Barcelona-Supercomputer Center under PRACE project number 2018194742.


\appendix

\section{Details simulations performed in cylindrical domain}

\begin{table*}[!h]
\caption{\label{table1} Direct numerical simulations in cylindrical domain for $Pr=4.38$ and $\Gamma=1$, matching the experiments by Zhong and Ahlers \cite{zho09,zho10c}. The columns from left to right indicate the $Ra$ number, the used numerical resolution in azimuthal, radial, and axial direction, the number of points in the thermal boundary layer (for the non-rotating case), the $1/Ro$ range considered in the simulations, and the number of considered cases.}
\begin{ruledtabular}
\begin{tabular}{cccccccccc}
$Ra$ & $N_{\theta}\times N_{r} \times N_{z}$ & $~N_{~\theta~{\rm BL}}$ & $1/Ro~\mathrm{[min-max]} $ & Cases \\[0.2cm]
$1.0\times10^{7}$	& $256 \times 96 \times 192$		& 14 			& $0-10$	& $14$ 	\\
$1.0\times10^{8}$	& $384 \times 129 \times 256$		& 15 			& $0-10$	& $14$ 	\\
$1.0\times10^{9}$	& $768 \times 256 \times 512$		& 22 			& $0-10$ 	& $14$ 	\\
$2.3\times10^{9}$	& $768 \times 256 \times 512$		& 19 			& $0-10$ 	& $13$ 	\\
$1.8\times10^{10}$	& $3072 \times 512 \times 1024$		& 25		 	& $0-2.5$ 	& $8$ 	\\
\end{tabular}
\end{ruledtabular}
\end{table*}

\bigskip

\section{Details simulations performed in periodic domain}
In this section, we give the numerical details of our simulations of the horizontally periodic domain. In each following table the columns from left to right show the rotation rate $1/Ro$, the aspect ratio $\Gamma$ (width/height) of the domain, the number of grid points (and the refinement factor) for the horizontal $N_x(m_x)$ and vertical direction $N_z(m_z)$, the Nusselt number $Nu$, the Reynolds number $Re$ defined by the rms value of velocity magnitude, and the viscous $\lambda_u$ and thermal $\lambda_\theta$ boundary layer thickness, respectively. The same width and discretization is used in both horizontal directions. Each table shows the simulation data for one combination of $Ra$ and $Pr$.

\begin{table*}[!h]
\caption{\label{tab:pr4ra7} $Pr=4.38$ and $Ra=1\times10^7$}
\begin{ruledtabular}
\begin{tabular}{cccccccc}
 $1/Ro$ & $\Gamma$ & $N_x(m_x)$ & $N_z(m_z)$ & $Nu$ & $Re$ & $\lambda_u/L$ & $\lambda_\theta/L$ \\[0.2cm]
 0.0 & 6.0 & 240(2) & 120(1) & 16.36 & 200.6 & $8.68\times10^{-2}$ & $2.86\times10^{-2}$ \\ 
 0.1 & 4.0 & 192(2) & 144(1) & 16.36 & 198.5 & $9.03\times10^{-2}$ & $2.84\times10^{-2}$ \\ 
 1.0 & 4.0 & 192(2) & 144(1) & 18.57 & 144.3 & $5.27\times10^{-2}$ & $2.55\times10^{-2}$ \\ 
 2.0 & 4.0 & 192(2) & 144(1) & 19.77 & 133.5 & $4.34\times10^{-2}$ & $2.58\times10^{-2}$ \\ 
 3.2 & 4.0 & 240(2) & 144(1) & 20.22 & 124.5 & $3.74\times10^{-2}$ & $2.77\times10^{-2}$ \\ 
 5.0 & 4.0 & 240(2) & 144(1) & 20.14 & 113.4 & $3.20\times10^{-2}$ & $2.87\times10^{-2}$ \\ 
 7.0 & 4.0 & 240(2) & 144(1) & 18.72 & 101.1 & $2.81\times10^{-2}$ & $3.01\times10^{-2}$ \\ 
 10.0 & 4.0 & 240(2) & 192(1) & 15.13 & 81.91 & $2.43\times10^{-2}$ & $3.50\times10^{-2}$ \\ 
 15.0 & 4.0 & 240(2) & 192(1) & 8.972 & 54.57 & $2.06\times10^{-2}$ & $5.07\times10^{-2}$ \\ 
\end{tabular}
\end{ruledtabular}
\end{table*}

\begin{table*}[!h]
\caption{\label{tab:pr4ra8} $Pr=4.38$ and $Ra=1\times10^8$}
\begin{ruledtabular}
\begin{tabular}{cccccccc}
 $1/Ro$ & $\Gamma$ & $N_x(m_x)$ & $N_z(m_z)$ & $Nu$ & $Re$ & $\lambda_u/L$ & $\lambda_\theta/L$ \\[0.2cm]
 0.0 & 5.0 & 360(3) & 144(2) & 31.21 & 643.9 & $7.07\times10^{-2}$ & $1.47\times10^{-2}$ \\ 
 0.1 & 5.0 & 360(3) & 144(2) & 31.43 & 628.8 & $7.09\times10^{-2}$ & $1.46\times10^{-2}$ \\ 
 1.0 & 3.0 & 288(2) & 144(2) & 34.73 & 410.3 & $3.31\times10^{-2}$ & $1.32\times10^{-2}$ \\ 
 2.0 & 3.0 & 288(3) & 192(1) & 36.44 & 373.7 & $2.64\times10^{-2}$ & $1.33\times10^{-2}$ \\ 
 3.2 & 3.0 & 288(3) & 192(1) & 38.00 & 355.5 & $2.21\times10^{-2}$ & $1.35\times10^{-2}$ \\ 
 5.0 & 3.0 & 288(3) & 192(1) & 38.80 & 330.6 & $1.86\times10^{-2}$ & $1.41\times10^{-2}$ \\ 
 7.0 & 3.0 & 288(3) & 192(1) & 38.51 & 307.5 & $1.62\times10^{-2}$ & $1.43\times10^{-2}$ \\ 
 10.0 & 3.0 & 288(3) & 192(1) & 36.09 & 275.7 & $1.39\times10^{-2}$ & $1.43\times10^{-2}$ \\ 
 15.0 & 3.0 & 288(3) & 192(1) & 28.36 & 220.2 & $1.16\times10^{-2}$ & $1.55\times10^{-2}$ \\ 
 20.0 & 2.0 & 288(2) & 240(1) & 19.14 & 163.7 & $1.01\times10^{-2}$ & $2.18\times10^{-2}$ \\ 
\end{tabular}
\end{ruledtabular}
\end{table*}

\begin{table*}[!h]
\caption{\label{tab:pr4ra58} $Pr=4.38$ and $Ra=5\times10^8$}
\begin{ruledtabular}
\begin{tabular}{cccccccc}
 $1/Ro$ & $\Gamma$ & $N_x(m_x)$ & $N_z(m_z)$ & $Nu$ & $Re$ & $\lambda_u/L$ & $\lambda_\theta/L$ \\[0.2cm]
 0.0 & 4.0 & 576(3) & 288(1) & 49.75 & 1437 & $5.66\times10^{-2}$ & $9.04\times10^{-3}$ \\ 
 0.1 & 4.0 & 576(3) & 288(1) & 50.42 & 1353 & $5.35\times10^{-2}$ & $8.93\times10^{-3}$ \\ 
 1.0 & 3.0 & 576(2) & 288(1) & 54.27 & 848.1 & $2.37\times10^{-2}$ & $8.06\times10^{-3}$ \\ 
 3.2 & 2.0 & 576(2) & 288(1) & 56.74 & 693.5 & $1.54\times10^{-2}$ & $8.28\times10^{-3}$ \\ 
 5.0 & 2.0 & 576(2) & 288(1) & 57.24 & 638.8 & $1.27\times10^{-2}$ & $8.63\times10^{-3}$ \\ 
 10.0 & 2.0 & 576(2) & 288(1) & 56.01 & 567.0 & $9.37\times10^{-3}$ & $8.73\times10^{-3}$ \\ 
 20.0 & 2.0 & 576(2) & 288(1) & 41.11 & 417.2 & $6.84\times10^{-3}$ & $8.93\times10^{-3}$ \\ 
 25.0 & 2.0 & 576(2) & 288(1) & 31.34 & 340.7 & $6.14\times10^{-3}$ & $1.12\times10^{-2}$ \\ 
\end{tabular}
\end{ruledtabular}
\end{table*}

\begin{table*}[!h]
\caption{\label{tab:pr4ra9} $Pr=4.38$ and $Ra=1\times10^9$}
\begin{ruledtabular}
\begin{tabular}{cccccccc}
 $1/Ro$ & $\Gamma$ & $N_x(m_x)$ & $N_z(m_z)$ & $Nu$ & $Re$ & $\lambda_u/L$ & $\lambda_\theta/L$ \\[0.2cm]
 0.0 & 3.0 & 512(3) & 256(2) & 62.11 & 1989 & $4.73\times10^{-2}$ & $7.13\times10^{-3}$ \\ 
 0.1 & 3.0 & 512(3) & 256(2) & 62.82 & 1886 & $4.62\times10^{-2}$ & $7.08\times10^{-3}$ \\ 
 1.0 & 2.0 & 384(3) & 256(2) & 66.43 & 1167 & $2.06\times10^{-2}$ & $6.47\times10^{-3}$ \\ 
 2.5 & 2.0 & 512(3) & 256(2) & 67.59 & 982.0 & $1.47\times10^{-2}$ & $6.50\times10^{-3}$ \\ 
 5.0 & 2.0 & 512(3) & 256(2) & 67.55 & 842.2 & $1.09\times10^{-2}$ & $6.83\times10^{-3}$ \\ 
 7.5 & 2.0 & 512(3) & 256(2) & 66.55 & 779.2 & $9.03\times10^{-3}$ & $7.01\times10^{-3}$ \\ 
 10.0 & 2.0 & 512(3) & 384(1) & 65.58 & 744.7 & $7.91\times10^{-3}$ & $7.07\times10^{-3}$ \\ 
 20.0 & 2.0 & 512(3) & 384(1) & 52.75 & 590.0 & $5.77\times10^{-3}$ & $7.07\times10^{-3}$ \\ 
 30.0 & 1.6 & 512(2) & 384(1) & 33.39 & 419.6 & $4.72\times10^{-3}$ & $9.82\times10^{-3}$ \\ 
\end{tabular}
\end{ruledtabular}
\end{table*}

\begin{table*}[!h]
\caption{\label{tab:pr4ra29} $Pr=4.38$ and $Ra=2.3\times10^9$}
\begin{ruledtabular}
\begin{tabular}{cccccccc}
 $1/Ro$ & $\Gamma$ & $N_x(m_x)$ & $N_z(m_z)$ & $Nu$ & $Re$ & $\lambda_u/L$ & $\lambda_\theta/L$ \\[0.2cm]
 0.0 & 3.0 & 648(3) & 288(2) & 80.42 & 2890 & $4.05\times10^{-2}$ & $5.45\times10^{-3}$ \\ 
 0.1 & 3.0 & 648(3) & 288(2) & 81.45 & 2694 & $3.94\times10^{-2}$ & $5.39\times10^{-3}$ \\ 
 0.4 & 3.0 & 648(3) & 288(2) & 83.84 & 2036 & $2.44\times10^{-2}$ & $5.18\times10^{-3}$ \\ 
 1.0 & 2.0 & 576(2) & 288(2) & 84.81 & 1731 & $1.83\times10^{-2}$ & $5.00\times10^{-3}$ \\ 
 1.6 & 1.6 & 432(3) & 288(3) & 85.05 & 1591 & $1.49\times10^{-2}$ & $4.95\times10^{-3}$ \\ 
 3.2 & 1.6 & 432(3) & 288(2) & 84.57 & 1350 & $1.11\times10^{-2}$ & $5.03\times10^{-3}$ \\ 
 7.5 & 1.2 & 360(3) & 360(2) & 79.23 & 1071 & $7.41\times10^{-3}$ & $5.37\times10^{-3}$ \\ 
 10.0 & 1.2 & 384(3) & 384(2) & 77.47 & 1005 & $6.48\times10^{-3}$ & $5.42\times10^{-3}$ \\ 
 20.0 & 1.2 & 384(3) & 384(2) & 67.06 & 860.3 & $4.71\times10^{-3}$ & $5.43\times10^{-3}$ \\ 
 40.0 & 1.2 & 384(3) & 384(1) & 30.75 & 484.4 & $3.37\times10^{-3}$ & $9.87\times10^{-3}$ \\ 
\end{tabular}
\end{ruledtabular}
\end{table*}

\begin{table*}[!h]
\caption{\label{tab:pr6ra7} $Pr=6.4$ and $Ra=1\times10^7$}
\begin{ruledtabular}
\begin{tabular}{cccccccc}
 $1/Ro$ & $\Gamma$ & $N_x(m_x)$ & $N_z(m_z)$ & $Nu$ & $Re$ & $\lambda_u/L$ & $\lambda_\theta/L$ \\[0.2cm]
 0.0 & 5.0 & 240(2) & 144(1) & 16.26 & 143.3 & $9.78\times10^{-2}$ & $2.91\times10^{-2}$ \\ 
 0.1 & 5.0 & 240(2) & 144(1) & 16.29 & 143.7 & $1.01\times10^{-1}$ & $2.91\times10^{-2}$ \\ 
 1.0 & 5.0 & 240(2) & 144(1) & 18.60 & 104.4 & $5.77\times10^{-2}$ & $2.59\times10^{-2}$ \\ 
 2.0 & 5.0 & 288(2) & 144(1) & 19.51 & 96.18 & $4.83\times10^{-2}$ & $2.69\times10^{-2}$ \\ 
 3.2 & 5.0 & 288(2) & 144(1) & 20.09 & 89.59 & $4.14\times10^{-2}$ & $2.77\times10^{-2}$ \\ 
 5.0 & 4.0 & 240(2) & 144(1) & 20.37 & 82.32 & $3.51\times10^{-2}$ & $2.82\times10^{-2}$ \\ 
 7.0 & 4.0 & 240(2) & 144(1) & 19.99 & 75.26 & $3.05\times10^{-2}$ & $2.83\times10^{-2}$ \\ 
 10.0 & 4.0 & 240(2) & 144(1) & 17.82 & 64.94 & $2.63\times10^{-2}$ & $3.09\times10^{-2}$ \\ 
 15.0 & 4.0 & 240(2) & 144(1) & 12.53 & 48.13 & $2.22\times10^{-2}$ & $4.24\times10^{-2}$ \\ 
 20.0 & 4.0 & 240(2) & 192(1) & 7.388 & 33.02 & $2.00\times10^{-2}$ & $5.81\times10^{-2}$ \\ 
\end{tabular}
\end{ruledtabular}
\end{table*}

\begin{table*}[!h]
\caption{\label{tab:pr6ra8} $Pr=6.4$ and $Ra=1\times10^8$}
\begin{ruledtabular}
\begin{tabular}{cccccccc}
 $1/Ro$ & $\Gamma$ & $N_x(m_x)$ & $N_z(m_z)$ & $Nu$ & $Re$ & $\lambda_u/L$ & $\lambda_\theta/L$ \\[0.2cm]
 0.0 & 4.0 & 288(3) & 144(2) & 31.01 & 479.9 & $7.66\times10^{-2}$ & $1.48\times10^{-2}$ \\ 
 0.1 & 4.0 & 288(3) & 144(2) & 31.34 & 459.2 & $7.29\times10^{-2}$ & $1.47\times10^{-2}$ \\ 
 1.0 & 3.0 & 288(3) & 192(2) & 34.70 & 299.8 & $3.67\times10^{-2}$ & $1.35\times10^{-2}$ \\ 
 2.0 & 3.0 & 288(3) & 192(2) & 36.89 & 276.9 & $2.92\times10^{-2}$ & $1.34\times10^{-2}$ \\ 
 5.0 & 3.0 & 288(3) & 192(2) & 39.82 & 243.2 & $2.03\times10^{-2}$ & $1.36\times10^{-2}$ \\ 
 7.0 & 3.0 & 360(3) & 240(1) & 39.90 & 225.7 & $1.76\times10^{-2}$ & $1.36\times10^{-2}$ \\ 
 10.0 & 3.0 & 360(3) & 240(1) & 38.81 & 205.2 & $1.51\times10^{-2}$ & $1.35\times10^{-2}$ \\ 
 20.0 & 2.0 & 240(3) & 240(1) & 25.60 & 137.6 & $1.11\times10^{-2}$ & $1.73\times10^{-2}$ \\ 
\end{tabular}
\end{ruledtabular}
\end{table*}

\begin{table*}[!h]
\caption{\label{tab:pr6ra9} $Pr=6.4$ and $Ra=1\times10^9$}
\begin{ruledtabular}
\begin{tabular}{cccccccc}
 $1/Ro$ & $\Gamma$ & $N_x(m_x)$ & $N_z(m_z)$ & $Nu$ & $Re$ & $\lambda_u/L$ & $\lambda_\theta/L$ \\[0.2cm]
 0.0 & 3.0 & 384(4) & 192(2) & 61.72 & 1512 & $5.43\times10^{-2}$ & $7.19\times10^{-3}$ \\ 
 0.1 & 3.0 & 384(4) & 192(2) & 62.51 & 1401 & $4.80\times10^{-2}$ & $7.11\times10^{-3}$ \\ 
 1.0 & 1.0 & 288(2) & 240(2) & 66.58 & 864.0 & $2.29\times10^{-2}$ & $6.63\times10^{-3}$ \\ 
 2.0 & 1.0 & 360(2) & 240(2) & 68.52 & 764.4 & $1.76\times10^{-2}$ & $6.64\times10^{-3}$ \\ 
 5.0 & 1.0 & 360(2) & 240(2) & 71.53 & 636.9 & $1.19\times10^{-2}$ & $6.78\times10^{-3}$ \\ 
 7.0 & 1.0 & 360(2) & 240(2) & 72.51 & 608.0 & $1.02\times10^{-2}$ & $6.77\times10^{-3}$ \\ 
 10.0 & 1.0 & 360(2) & 240(2) & 71.86 & 561.9 & $8.60\times10^{-3}$ & $6.61\times10^{-3}$ \\ 
 20.0 & 1.0 & 384(2) & 240(2) & 62.87 & 461.5 & $6.27\times10^{-3}$ & $6.42\times10^{-3}$ \\ 
 40.0 & 1.0 & 288(3) & 288(1) & 27.56 & 249.3 & $4.53\times10^{-3}$ & $1.24\times10^{-2}$ \\ 
\end{tabular}
\end{ruledtabular}
\end{table*}

\begin{table*}[!h]
\caption{\label{tab:pr25ra7} $Pr=25$ and $Ra=1\times10^7$}
\begin{ruledtabular}
\begin{tabular}{cccccccc}
 $1/Ro$ & $\Gamma$ & $N_x(m_x)$ & $N_z(m_z)$ & $Nu$ & $Re$ & $\lambda_u/L$ & $\lambda_\theta/L$ \\[0.2cm]
 0.0 & 6.0 & 256(4) & 192(1) & 16.23 & 41.11 & $1.42\times10^{-1}$ & $3.13\times10^{-2}$ \\ 
 0.1 & 4.0 & 192(4) & 128(2) & 16.36 & 42.14 & $1.38\times10^{-1}$ & $3.08\times10^{-2}$ \\ 
 1.0 & 4.0 & 192(4) & 128(2) & 17.92 & 31.54 & $8.17\times10^{-2}$ & $2.85\times10^{-2}$ \\ 
 4.0 & 3.0 & 192(4) & 256(1) & 19.50 & 25.11 & $5.13\times10^{-2}$ & $2.79\times10^{-2}$ \\ 
 10.0 & 3.0 & 192(4) & 256(1) & 20.20 & 21.41 & $3.54\times10^{-2}$ & $2.74\times10^{-2}$ \\ 
 15.0 & 3.0 & 192(4) & 256(1) & 19.83 & 19.48 & $2.98\times10^{-2}$ & $2.75\times10^{-2}$ \\ 
 20.0 & 3.0 & 192(4) & 256(1) & 18.43 & 17.78 & $2.64\times10^{-2}$ & $2.98\times10^{-2}$ \\ 
 30.0 & 3.0 & 192(4) & 256(1) & 11.75 & 12.68 & $2.23\times10^{-2}$ & $4.94\times10^{-2}$ \\ 
\end{tabular}
\end{ruledtabular}
\end{table*}

\begin{table*}[!h]
\caption{\label{tab:pr25ra8} $Pr=25$ and $Ra=1\times10^8$}
\begin{ruledtabular}
\begin{tabular}{cccccccc}
 $1/Ro$ & $\Gamma$ & $N_x(m_x)$ & $N_z(m_z)$ & $Nu$ & $Re$ & $\lambda_u/L$ & $\lambda_\theta/L$ \\[0.2cm]
 0.0 & 5.0 & 384(4) & 192(2) & 31.05 & 145.3 & $1.00\times10^{-1}$ & $1.53\times10^{-2}$ \\ 
 0.01 & 5.0 & 384(4) & 192(2) & 30.95 & 146.3 & $1.05\times10^{-1}$ & $1.54\times10^{-2}$ \\ 
 0.1 & 4.0 & 288(4) & 192(2) & 31.36 & 140.3 & $9.28\times10^{-2}$ & $1.51\times10^{-2}$ \\ 
 1.0 & 4.0 & 288(4) & 192(2) & 34.31 & 95.40 & $5.22\times10^{-2}$ & $1.45\times10^{-2}$ \\ 
 10.0 & 3.0 & 288(5) & 192(2) & 41.10 & 63.68 & $2.06\times10^{-2}$ & $1.26\times10^{-2}$ \\ 
 15.0 & 3.0 & 288(5) & 288(1) & 42.04 & 59.43 & $1.72\times10^{-2}$ & $1.24\times10^{-2}$ \\ 
 20.0 & 2.0 & 288(4) & 288(1) & 41.73 & 55.26 & $1.51\times10^{-2}$ & $1.23\times10^{-2}$ \\ 
 40.0 & 2.0 & 240(4) & 288(1) & 28.82 & 38.59 & $1.11\times10^{-2}$ & $1.77\times10^{-2}$ \\ 
 60.0 & 2.0 & 240(4) & 288(1) & 13.91 & 24.12 & $9.28\times10^{-3}$ & $3.06\times10^{-2}$ \\ 
\end{tabular}
\end{ruledtabular}
\end{table*}

\begin{table*}[!h]
\caption{\label{tab:pr25ra9} $Pr=25$ and $Ra=1\times10^9$}
\begin{ruledtabular}
\begin{tabular}{cccccccc}
 $1/Ro$ & $\Gamma$ & $N_x(m_x)$ & $N_z(m_z)$ & $Nu$ & $Re$ & $\lambda_u/L$ & $\lambda_\theta/L$ \\[0.2cm]
 0.0 & 3.0 & 384(6) & 288(3) & 60.19 & 521.0 & $7.08\times10^{-2}$ & $7.46\times10^{-3}$ \\ 
 0.1 & 2.0 & 384(4) & 288(3) & 61.42 & 451.2 & $6.03\times10^{-2}$ & $7.35\times10^{-3}$ \\ 
 1.0 & 2.0 & 384(4) & 288(3) & 66.29 & 285.3 & $3.21\times10^{-2}$ & $7.20\times10^{-3}$ \\ 
 10.0 & 1.0 & 288(4) & 384(2) & 81.20 & 189.1 & $1.19\times10^{-2}$ & $6.11\times10^{-3}$ \\ 
 15.0 & 1.0 & 288(4) & 384(2) & 84.23 & 175.9 & $9.87\times10^{-3}$ & $5.80\times10^{-3}$ \\ 
 20.0 & 1.0 & 288(4) & 384(2) & 84.67 & 165.1 & $8.64\times10^{-3}$ & $5.67\times10^{-3}$ \\ 
 30.0 & 1.0 & 288(4) & 384(2) & 79.03 & 141.4 & $7.13\times10^{-3}$ & $5.69\times10^{-3}$ \\ 
 40.0 & 1.0 & 288(4) & 576(1) & 68.71 & 121.2 & $6.21\times10^{-3}$ & $6.02\times10^{-3}$ \\ 
 60.0 & 1.0 & 288(4) & 576(1) & 49.06 & 92.58 & $5.14\times10^{-3}$ & $8.35\times10^{-3}$ \\ 
 80.0 & 1.0 & 288(4) & 576(1) & 31.59 & 69.74 & $4.51\times10^{-3}$ & $1.24\times10^{-2}$ \\ 
\end{tabular}
\end{ruledtabular}
\end{table*}

\begin{table*}[!h]
\caption{\label{tab:pr100ra7} $Pr=100$ and $Ra=1\times10^7$}
\begin{ruledtabular}
\begin{tabular}{cccccccc}
 $1/Ro$ & $\Gamma$ & $N_x(m_x)$ & $N_z(m_z)$ & $Nu$ & $Re$ & $\lambda_u/L$ & $\lambda_\theta/L$ \\[0.2cm]
 0.0 & 5.0 & 192(4) & 192(1) & 16.67 & 10.54 & $1.62\times10^{-1}$ & $3.18\times10^{-2}$ \\ 
 0.1 & 4.0 & 192(4) & 192(1) & 16.61 & 11.13 & $1.56\times10^{-1}$ & $3.18\times10^{-2}$ \\ 
 1.0 & 4.0 & 192(4) & 192(1) & 17.21 & 8.833 & $1.07\times10^{-1}$ & $3.08\times10^{-2}$ \\ 
 10.0 & 4.0 & 240(4) & 192(1) & 19.63 & 6.056 & $4.72\times10^{-2}$ & $2.75\times10^{-2}$ \\ 
 20.0 & 4.0 & 288(4) & 192(1) & 19.75 & 5.551 & $3.58\times10^{-2}$ & $2.89\times10^{-2}$ \\ 
 30.0 & 3.0 & 240(4) & 192(1) & 19.02 & 4.979 & $3.00\times10^{-2}$ & $2.93\times10^{-2}$ \\ 
 40.0 & 3.0 & 240(4) & 240(1) & 16.74 & 4.323 & $2.65\times10^{-2}$ & $3.15\times10^{-2}$ \\ 
 60.0 & 2.0 & 288(2) & 240(1) & 9.823 & 2.927 & $2.22\times10^{-2}$ & $6.29\times10^{-2}$ \\ 
\end{tabular}
\end{ruledtabular}
\end{table*}

\begin{table*}[!h]
\caption{\label{tab:pr100ra7} $Pr=100$ and $Ra=1\times10^8$}
\begin{ruledtabular}
\begin{tabular}{cccccccc}
 $1/Ro$ & $\Gamma$ & $N_x(m_x)$ & $N_z(m_z)$ & $Nu$ & $Re$ & $\lambda_u/L$ & $\lambda_\theta/L$ \\[0.2cm]
 0.0 & 5.0 & 384(4) & 192(2) & 31.44 & 41.03 & $1.48\times10^{-1}$ & $1.60\times10^{-2}$ \\ 
 0.01 & 3.0 & 288(3) & 192(2) & 31.50 & 43.04 & $1.41\times10^{-1}$ & $1.60\times10^{-2}$ \\ 
 0.1 & 3.0 & 288(4) & 192(2) & 31.66 & 40.11 & $1.29\times10^{-1}$ & $1.59\times10^{-2}$ \\ 
 1.0 & 3.0 & 288(4) & 216(2) & 33.40 & 27.83 & $7.29\times10^{-2}$ & $1.55\times10^{-2}$ \\ 
 10.0 & 2.0 & 288(4) & 216(2) & 38.90 & 17.42 & $2.78\times10^{-2}$ & $1.30\times10^{-2}$ \\ 
 20.0 & 2.0 & 288(5) & 216(2) & 41.91 & 16.77 & $2.08\times10^{-2}$ & $1.26\times10^{-2}$ \\ 
 30.0 & 2.0 & 288(5) & 216(2) & 43.07 & 16.21 & $1.74\times10^{-2}$ & $1.26\times10^{-2}$ \\ 
 35.0 & 2.0 & 360(5) & 240(2) & 43.57 & 15.89 & $1.62\times10^{-2}$ & $1.25\times10^{-2}$ \\ 
 40.0 & 2.0 & 360(5) & 240(2) & 43.57 & 15.41 & $1.53\times10^{-2}$ & $1.24\times10^{-2}$ \\ 
 60.0 & 1.2 & 240(4) & 288(1) & 35.80 & 12.24 & $1.26\times10^{-2}$ & $1.35\times10^{-2}$ \\ 
100.0 & 1.0 & 240(3) & 240(1) & 18.47 & 7.79 & $9.99\times10^{-3}$ & $2.46\times10^{-2}$ \\ 
\end{tabular}
\end{ruledtabular}
\end{table*}

\begin{table*}[!h]
\caption{\label{tab:pr100ra7} $Pr=100$ and $Ra=1\times10^9$}
\begin{ruledtabular}
\begin{tabular}{cccccccc}
 $1/Ro$ & $\Gamma$ & $N_x(m_x)$ & $N_z(m_z)$ & $Nu$ & $Re$ & $\lambda_u/L$ & $\lambda_\theta/L$ \\[0.2cm]
 0.0 & 3.0 & 480(6) & 384(2) & 60.82 & 164.4 & $1.01\times10^{-1}$ & $7.60\times10^{-3}$ \\ 
 0.1 & 2.0 & 384(4) & 384(2) & 61.82 & 137.1 & $7.96\times10^{-2}$ & $7.54\times10^{-3}$ \\ 
 1.0 & 2.0 & 384(5) & 384(2) & 64.49 & 90.93 & $4.61\times10^{-2}$ & $7.64\times10^{-3}$ \\ 
 10.0 & 1.0 & 384(4) & 384(2) & 78.28 & 52.76 & $1.64\times10^{-2}$ & $6.28\times10^{-3}$ \\ 
 20.0 & 1.0 & 384(4) & 384(2) & 85.38 & 49.40 & $1.19\times10^{-2}$ & $5.77\times10^{-3}$ \\ 
 30.0 & 1.0 & 432(4) & 648(1) & 89.82 & 47.24 & $9.92\times10^{-3}$ & $5.63\times10^{-3}$ \\ 
 36.0 & 1.0 & 432(4) & 648(1) & 92.28 & 46.48 & $9.11\times10^{-3}$ & $5.48\times10^{-3}$ \\ 
 40.0 & 1.0 & 432(4) & 648(1) & 92.06 & 45.12 & $8.66\times10^{-3}$ & $5.51\times10^{-3}$ \\ 
 80.0 & 1.0 & 432(4) & 648(1) & 79.67 & 34.79 & $6.24\times10^{-3}$ & $6.17\times10^{-3}$ \\ 
120.0 & 1.0 & 432(4) & 648(1) & 52.12 & 25.73 & $5.15\times10^{-3}$ & $7.42\times10^{-3}$ \\ 
\end{tabular}
\end{ruledtabular}
\end{table*}

\FloatBarrier

\end{document}